# Acoustic confinement and Stimulated Brillouin Scattering in integrated optical waveguides

Christopher G. Poulton,[1,2,3,*]  Ravi Pant,[1,3] and  Benjamin J. Eggleton[1,3]

[1]*Centre for Ultrahigh bandwidth Devices for Optical Systems (CUDOS)*
[2]*School of Mathematical Sciences, University of Technology Sydney, NSW 2007, Australia*
[3]*School of Physics, University of Sydney, NSW 2006, Australia*

We examine the effect of acoustic mode confinement on Stimulated Brillouin Scattering in optical waveguides that consist of a guiding core embedded in a solid substrate. We find that SBS can arise due to coupling to acoustic modes in three different regimes. First, the acoustic modes may be guided by total internal reflection; in this case the SBS gain depends directly on the degree of confinement of the acoustic mode in the core, which is in turn determined by the acoustic V-parameter. Second, the acoustic modes may be leaky, but may nevertheless have a sufficiently long lifetime to have a large effect on the SBS gain; the lifetime of acoustic modes in this regime depends not only on the contrast in acoustic properties between the core and the cladding, but is also highly dependent on the waveguide dimensions. Finally SBS may occur due to coupling to free modes, which exist even in the absence of acoustic confinement; we find that the cumulative effect of coupling to these non-confined modes results in significant SBS gain. We show how the different acoustic properties of core and cladding lead to these different regimes, and discuss the feasibility of SBS experiments using different material systems.

*OCIS codes:*    (190.0190) Nonlinear optics; (190.2640) Stimulated scattering, modulation, etc.

http://dx.doi.org/10.1364/XX.99.099999

## 1. Introduction

Stimulated Brillouin Scattering (SBS) occurs when light interacts coherently with acoustic waves via a combination of electrostriction and the photo-elastic effect [1]. Although SBS is traditionally associated with very long lengths of optical fibres, a number of recent experiments have demonstrated and exploited SBS in much shorter devices, including microtoroids, microspheres, and wedge-profiled microdisks [2–4], integrated optical waveguides [5] and multi-structured optical fibres [6]. These demonstrations open the door to a number of important on-chip applications, such as tunable microwave photonic filters [7, 8], ultra-sensitive frequency sensors [9, 10], compact generators of slow and fast light [11, 12] and all-optical isolation [13, 14]. In on-chip devices, SBS also provides a means of accessing the quantum optomechanical regime [15]. For sufficiently small waveguide dimensions it has recently been predicted [16] that radiation pressure will lead to significant enhancement of the SBS gain, thereby enabling SBS to be used in CMOS-compatible platforms such as Silicon [17].

The material properties of the SBS gain medium play an important role in the ability to induce and control SBS on very short scales. It is well-known that the SBS gain $g_B$ is proportional to $n^8$, where $n$ is the refractive index of the waveguiding material; the gain can therefore be increased by using high-index optically transparent materials such as chalcogenide glasses, which have $n \sim 2.4 - 2.8$ [18]. One can further improve the interaction by using materials with a high electrostrictive constant (or equivalently, a high photo-elastic constant). Embedding such a material in a lower-index material ensures that the light is confined to the core and so the optical intensity remains high. However, it is perhaps less well appreciated that the *acoustic* configuration of the waveguide is also very important in the SBS interaction: if both optical and acoustic waves are confined within the same volume then the SBS process will be driven efficiently; on the other hand, a low opto-acoustic overlap will lead to a corresponding reduction in the SBS gain, and this may lead to no effect being observed at all. The configuration of acoustic properties is therefore a key factor that must be understood for SBS to be successfully used in on-chip applications.

In Fig.1 we illustrate several waveguide configurations that have either been used or proposed for SBS experiments. These structures can be categorized as *suspended* structures, which include simple suspended nanowires (a), membranes (b), and suspended-core multi-structured fibres (c), and *embedded* structures, which include buried waveguides (d), on-substrate nanowires (e) and rib waveguides (f). Suspended struc-tures confine the acoustic mode using the high acoustic contrast of a solid-air interface, which traps the vibration in the solid material. The principle advantage of this approach is that the acoustic guidance mechanism is well understood and can be applied to any material. However, suspended solid-air structures are potentially limited in length by structural weaknesses of the material, thereby reducing their on-chip applicability. Embedded waveguides, which make use of the material properties of core and cladding to trap light and sound in the same volume, are an attractive alternative to on-chip generation of SBS.

---

*[*] Chris.Poulton@uts.edu.au

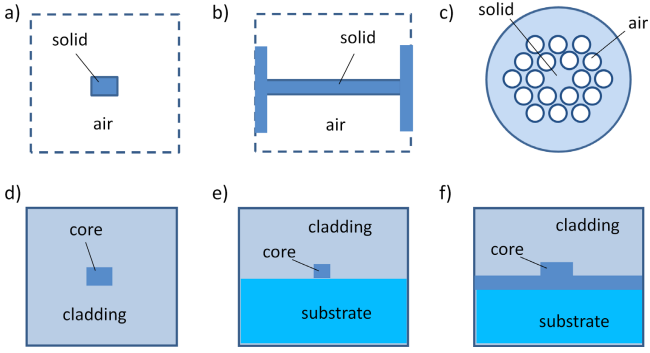

Fig. 1. Different waveguide geometries proposed for SBS experiments; a) Simple suspended nanowire; b) suspended membrane c); suspended-core multistructured optical fibre; d) buried waveguide; e) optical nanowire; f) rib waveguide.

The feasibility of using embedded waveguides has been demonstrated in a number of recent SBS experiments [5, 8, 12], which used chalcogenide rib waveguides placed on a silica substrate and with a cover material consisting either of silica [5] or IPG polymer [8, 12]. In these studies it was shown using simulations that the acoustic modes are confined to the waveguide core, thereby enabling large SBS gain. However the mechanisms resulting in acoustic confinement for these and other embedded waveguides have not yet been fully explained. A better understanding of the link between acoustic material properties and SBS is required in order to explore the effect in other on-chip systems.

In this paper we investigate the dependence of SBS gain on the acoustic properties of embedded waveguides. We find that the SBS gain depends on coupling to acoustic modes that exist in one of three regimes. In addition to acoustic modes that are guided using total internal reflection (TIR), there exist a class of leaky acoustic waves that are the acoustic analogue of hollow-core waveguides, and that can also couple to the optical field. Although these modes are not strictly confined, they are nevertheless localized for sufficiently long times to have a significant effect on the SBS gain. The lifetime of these leaky acoustic modes, and therefore their effect on the gain, scales in a simple way on the acoustic properties of the core and cladding material, and we show how this mode lifetime will behave for several technologically important material systems. We also find that, in the complete absence of any acoustic guided modes, the coupling to the spectrum of free acoustic modes can have a substantial effect on the gain, which manifests as an asymmetric gain peak occurring at frequencies above the SBS frequency shift in the bulk material.

## 2. Stimulated Brillouin scattering and the opto-acoustic overlap

We consider the interaction in an optical waveguide between a pump beam at frequency $\omega_p$ and a counter-propagating signal at frequency $\omega$, which is downshifted from the pump and lies close to the Brillouin Stokes line for the guiding material. For the sake of simplicity we restrict our investigation to "backward-SBS", in which the pump and Stokes signal are counter-propagating. We also restrict our study to waveguides in which the SBS interaction is dominated by longitudinal modes, and those in which the effects of radiation pressure can be neglected [16]. Finally, we also assume that both the pump and signal occupy a single optical mode, which we take to be the fundamental mode with dominant field component aligned in the $x$-direction. For the SBS interaction to occur the waveguide structure must also support acoustic modes: these we label with the subscript $m$, so that each acoustic mode has angular frequency $\Omega_m$ and propagation constant $q_m$ (Fig.2b). If we denote the propagation constants of the signal and pump by $\beta$ and $\beta_p$ respectively, then at the SBS gain peak we have the frequency/phase relations $\Omega_m = \omega_p - \omega$ and $q = \beta_p - \beta \approx 2\beta_p$.

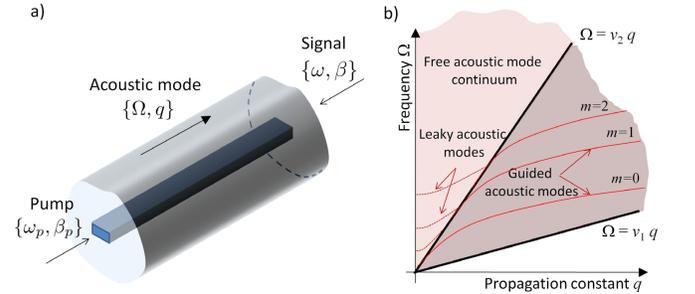

Fig. 2. a) Schematic of the SBS interaction between pump, signal and acoustic wave in a waveguide; b) Dispersion diagram illustrating the different regimes of acoustic modes.

When the power in the pump is much larger than that of the signal, the gain in the signal power due to the SBS interaction with the $m^{\text{th}}$ acoustic mode is

$$G_m = g_m^I I_p L \qquad (1)$$

where $I_p$ is the pump intensity, $L$ is the waveguide length, and $g_m^I$, in units of mW$^{-1}$, is given by [19]:

$$g_m^I = \frac{16\pi^3 n_1^8 p_{12}^2}{c \lambda_p^3 \rho_0 \Omega_m \Gamma_m} \frac{(\Gamma_m/2)^2}{(\omega - \omega_p + \Omega_m)^2 + (\Gamma_m/2)^2} \eta_m . \quad (2)$$

Here $\rho_0$ is the material density, $p_{12}$ is the elasto-optic coefficient, $n_1$ is the refractive index of the waveguide core and $\Gamma_m/2\pi$ is the Brillouin linewidth of the gain material. The quantity $\eta_m$ depends on the overlap between the optical intensity $||\mathbf{E}||^2$ and the acoustic mode field $\tilde{\rho}_m$:

$$\eta_m = \frac{\left|\int ||\mathbf{E}||^2 \tilde{\rho}_m \mathrm{d}A\right|^2}{\int ||\mathbf{E}||^2 \mathrm{d}A \int |\tilde{\rho}_m|^2 \mathrm{d}A} \, , \tag{3}$$

in which the integrals extend over the entire waveguide cross-section. $\eta_m$ is a dimensionless quantity, lying in the range $0 \leq \eta_m \leq 1$, that directly determines the efficiency of the Brillouin interaction, and is therefore suitable for directly comparing the SBS properties of different waveguides, as well as of different modes of the same waveguide. We refer to $\eta_m$ as the *opto-acoustic overlap* of mode $m$. The main objective of this investigation is to show how this quantity depends on interaction with the different types of acoustic modes that can exist within the waveguide.

In order to compute the SBS coupling, we must first calculate the frequencies and fields of the acoustic modes of the structure. The governing equation for a longitudinal acoustic mode in a waveguide aligned along the $z$ axis, with propagation constant $q$ and angular frequency $\Omega$, is

$$\nabla_\perp^2 \tilde{\rho} + \left(\frac{\Omega^2}{v_{1,2}^2} - q^2\right) \tilde{\rho} = 0 \, , \tag{4}$$

where $\nabla_\perp = \partial_x^2 + \partial_y^2$ is the transverse Laplacian operator and $v_j$ is the longitudinal sound velocity of the core (medium 1) or cladding (medium 2). In addition the following boundary conditions, arising from the continuity of the displacement and normal components of stress, must be fulfilled on the core-cladding interface:

$$\tilde{\rho}|_{\partial C^-} = \tilde{\rho}|_{\partial C^+} \quad , \quad \mu_1 \left.\frac{\partial \tilde{\rho}}{\partial n}\right|_{\partial C^-} = \mu_2 \left.\frac{\partial \tilde{\rho}}{\partial n}\right|_{\partial C^+} \tag{5}$$

where $\mu_{1,2}$ is the shear modulus of the core or cladding, and $\partial C$ represents the interface between the core and cladding.

The longitudinal modes arising from the solution of Eq. 4 manifest in two classes (see Fig. 1b): *guided modes*, in which the acoustic field is confined to the waveguide core using total internal reflection, and *leaky modes*, in which a wave near grazing incidence is confined by strong reflections at the core-cladding interface. In order to obtain a well-posed eigenvalue problem for the mode, appropriate boundary conditions at infinity must be specified; for guided modes the field $\tilde{\rho}$ must vanish far from the core, whereas for leaky modes the field is permitted to diverge provided that an outgoing wave condition is satisfied.

## 3. SBS coupling to guided acoustic modes

We begin by examining the solutions for the canonical case of SBS coupling to the guided modes of a step-index circular waveguide, which can be computed analytically. We consider a circular cylinder of radius $a$ and acoustic parameters $\{v_1, \mu_1\}$ embedded in an infinite matrix with acoustic parameters $\{v_2, \mu_2\}$ (see Fig.3a). The acoustic modes of this structure can be computed by expanding $\tilde{\rho}$ in terms of cylindrical functions and applying the boundary conditions (5) at $r = a$. This results in a simple transcendental equation that can be solved to find the frequency $\Omega_m$ for a given propagation constant $q$. An important quantity for determining the character of the acoustic solutions is the acoustic $V$-parameter $V_{\mathrm{ac}}$; for a circular acoustic guide this is

$$V_{\mathrm{ac}} = \Omega a \sqrt{\frac{1}{v_1^2} - \frac{1}{v_2^2}} \tag{6}$$

An inspection of Eq. (4) confirms that $V_{\mathrm{ac}}$ plays the same role in acoustics as it does in optical fibres, with the sound velocity playing the role of $c/n$. A sufficient condition for the existence of a guided acoustic mode is that $V_{\mathrm{ac}} > 0$, so that $v_1 < v_2$. In a direct analogy with the optical $V$-parameter, $V_{\mathrm{ac}}$ is a normalized measure of the contrast in acoustic velocity between the core and the cladding.

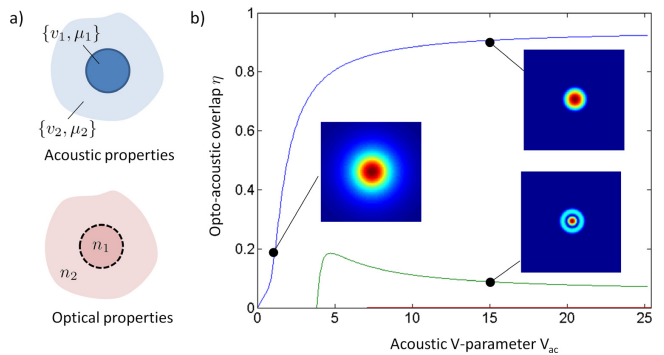

Fig. 3. a) Waveguide cross-section, showing optical and acoustic properties; b) Opto-acoustic overlap for acoustically guiding modes ($v_1 < v_2$) as a function of acoustic contrast, as represented by the acoustic V-number in Eq. (6).

In Fig. 3(a) we show how the opto-acoustic overlap $\eta_m$ changes with $V_{\mathrm{ac}}$ for the first three circularly symmetric acoustic modes, which are the modes that have the correct symmetry to couple to the fundamental optical mode. In this computation we have kept the optical properties fixed, with $n_1 = 2.37$ and $n_2 = 1.44$; we have chosen these parameters, which correspond to the somewhat artificial situation of a chalcogenide core in a silicon cladding, to illustrate the effect of the acoustic confinement as dramatically as possible. In addition, we consider coupling to the fundamental optical mode only. It can be seen that the $m = 0$ (fundamental) acoustic mode has a far larger overlap than the higher-order modes. As each acoustic mode approaches cutoff the acoustic field $\tilde{\rho}$ becomes less well confined to the core, thereby reducing the overlap in Eq. (3). Near to cutoff a maximum may occur for the higher-order modes. This is due to the acoustic field more closely overlapping with the fundamental optical mode as it spreads out from the core.

### 4. SBS coupling to leaky acoustic modes

The solution of the eigenvalue problem Eq. (4) - (5) also leads to solutions that have non-real values of $\Omega$. These correspond to leaky, or anti-guided modes, which are the acoustic analogue of the modes of hollow-core optical waveguides [20]. The leaky modes are confined due to the strong reflection of waves at grazing incidence to the core-cladding interface; unlike TIR-guided modes they are capable of carrying energy away from the waveguide. An important point is that these modes must still remain phase-matched to the two counter-propagating optical modes; for this reason it is convenient to describe these modes as having a real value of propagation constant $q$ and a complex acoustic frequency $\Omega$. This choice is contrary to the usual treatment of leaky modes, and leads to a description where the leaky modes decay in time, rather than space. However, the choice of complex $\Omega$ is convenient for the treatment of SBS interactions, in which the acoustic mode lifetime $\tau = 1/\text{Im}[\Omega]$ determines the SBS linewidth via the linewidth parameter $\Gamma = 1/\pi\tau$. If the mode lifetime is less than the phonon lifetime in the material then the SBS linewidth will be increased beyond that of the bulk material by the mode leakage, resulting in a reduction of SBS peak gain in accordance with Eq. (2). For acoustic mode lifetimes greater than the phonon lifetime, the SBS gain will be unaffected by the mode leakage; the SBS linewidth can then be expected to be similar to that of a non-leaky mode. The SBS peak gain in this case will be determined by the opto-acoustic overlap $\eta$.

Figure 4(a) shows the opto-acoustic overlap for the first three leaky modes as a function of acoustic contrast, as measured by the real quantity $iV_\text{ac}$. As in the guided case, the fundamental mode dominates the interaction, and the overlap of all modes decreases rapidly for small acoustic contrasts. When compared to the guided mode case in Fig. 3, it can be seen that the overlap with the leaky modes is smaller; this is because the leaky modes are in general less well confined to the core.

We can obtain an analytic approximation for the mode lifetime of the leaky acoustic modes using a perturbation expansion to the solution of Eqs. (4) - (5) in cylindrical coordinates, and further assuming that the guided mode makes a small angle of incidence upon striking the cladding boundary from the core (see Appendix). We obtain the following expression for the mode lifetime of the fundamental leaky mode of a cylindrical waveguide of radius $a$:

$$\tau \sim \frac{\Omega^2 a^3 \mu_2}{v_1^2 j_{00}^2 \mu_1} \sqrt{\frac{1}{v_2^2} - \frac{1}{v_1^2}} \qquad (7)$$

where $j_{00} \approx 2.405$ is the first zero of the Bessel function

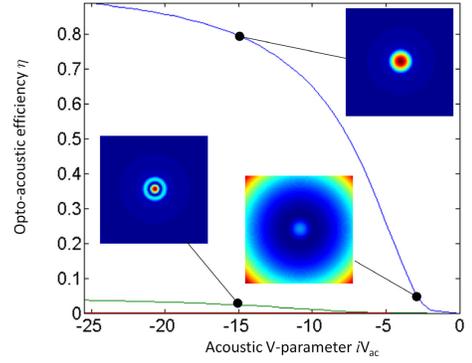

Fig. 4. Opto-acoustic overlap for leaky acoustic modes ($v_1 > v_2$) as a function of acoustic contrast, as represented by the imaginary part of the acoustic V-number.

$J_0(z)$. The rate of acoustic energy lost by the core due to mode leakage ($1/\tau$) scales inversely with the cube of the waveguide radius and is inversely proportional to the acoustic V-parameter $V_\text{ac}$. For a material systems of chalcogenide glass (n = 2.37, $v_1$ = 2600 ms$^{-1}$, $\mu_1$ = 6.34 GPa), embedded in a polymer (n = 1.44, $v_2$ = 1500 ms$^{-1}$, $\mu_2$ = 0.8 GPa), the lifetime of the acoustic mode is $\tau = 67ns$ for a 5 $\mu$m diameter waveguide, well above both the acoustic lifetime of the material ($\sim$ 10 ns) and the lifetime of quasi-CW pulses used in SBS experiments. In Fig. 5 we show the acoustic leakage lifetime, as computed from Eq. (7), for four different core/cladding systems for which $v_1 < v_2$ and so which support no guided acoustic modes. It can be seen that radii $> 2\mu$m are required to obtain significant $> 10$ns phonon lifetimes for each of these material systems. These results suggest that the small nanowire waveguides favored in many nonlinear experiments are likely to lead to short phonon lifetimes, thereby substantially reducing the SBS gain.

### 5. SBS coupling to the free acoustic mode spectrum

Even in the absence of confined modes it is possible to obtain significant SBS backscattering. This coupling is mediated via free-space, or radiation, modes, which exist in a continuum at acoustic frequencies $\Omega > v_2 q$. The phase matching diagram for this process is shown in Fig. 6: it can be seen that there exists a set of free modes for each point along the line $\Omega_a = c/n(\beta_p + \beta)$. The modes occurring at each of these values of $\Omega_a$ can be represented as an orthonormal set with labels $m = 0, 1, 2, ...$, with each value of $m$ corresponding to a different azimuthal order [21]. The gain at acoustic frequency $\Omega$ is then

$$g^I(\Omega) = \sum_{m=0}^{\infty} \int d\Omega_a g_m^I(\Omega, \Omega_a) \qquad (8)$$

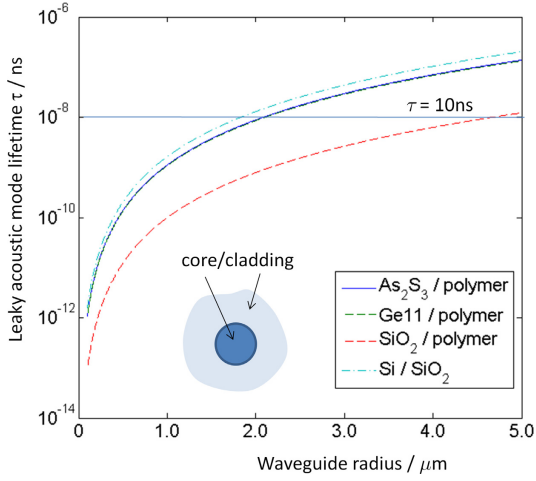

Fig. 5. Acoustic mode lifetimes for a cylindrical waveguide as a function of waveguide radius for four different core / cladding materials, computed from Eq. (7). The horizontal line at 10ns represents the phonon lifetime in chalcogenide glass.

For this simulation we again use a circular core with $n_1 = 2.37$ surrounded by a solid with index $n_2 = 1.44$. To demonstrate the effect of the coupling to the radiation field we choose the acoustic properties of the surrounding material to exactly match the acoustic properties of the chalcogenide core, however we specify that the SBS gain occurs only within the core itself. For acoustic frequencies beyond that of the bulk material ($\Omega_a > qv_2 = 2\pi \times 7.9 \times 10^9$ GHz) there exists a continuum of free modes (shaded region in Fig. 7). The coupling to these modes decreases as the acoustic modes move deeper into the free-mode region because the transverse wavenumber becomes larger, leading to a greater number of nodes within the overlap region. Although the individual contribution of each mode in the continuum is vanishingly small, the integrated gain is non-zero and manifests in a characteristic asymmetric SBS response. This asymmetry arises from the sudden appearance of the continuum modes at $\Omega_a = qv_2$: for frequencies below the continuum edge ($\Omega_a < qv_2$) the cumulative effect of the Lorentzian tails from the continuum modes results in a fall-off that is approximately equal to the Brillouin linewidth.

with the gain $g_m^I(\Omega, \Omega_a)$ for the $m^{\text{th}}$ acoustic mode at frequency $\Omega_a$ is given by

$$g_m^I(\Omega, \Omega_a) = \frac{16\pi^3 n_1^8 p_{12}^2}{c\lambda_p^3 \rho_0 \Omega_m \Gamma_m} \frac{(\Gamma_m/2)^2}{(\omega - \omega_p + \Omega_m)^2 + (\Gamma_m/2)^2}$$
$$\times \frac{\left|\int ||\mathbf{E}||^2 \tilde{\rho}_m^{\text{rad}}(\Omega_a) dA\right|^2}{\int ||\mathbf{E}||^2 dA} \quad . \tag{9}$$

Here the free modes $\tilde{\rho}_m^{\text{rad}}(\Omega_a)$ obey the differential equation (4) together with the outgoing wave condition at infinity. We note that the equation (9) differs from (2) and (3) in the normalisation term in the denominator, because the free acoustic modes must be normalized using a Dirac delta-function. The free modes are not bound to the waveguide, propagating in the cladding material with angle $\theta = \cos^{-1} \Omega_a/qv_2$ (see Fig. 6a).

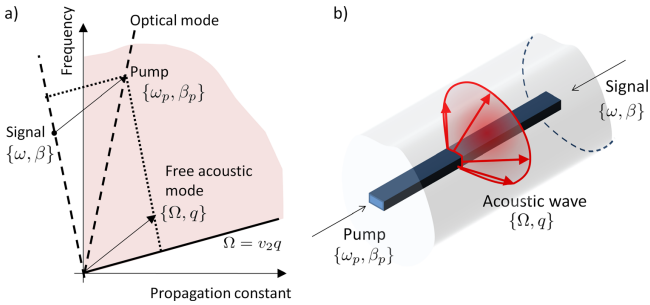

Fig. 6. a) Dispersion diagram showing the phase matching for SBS coupling via acoustic radiation modes; (b) Illustration of coupling to an acoustic radiation mode, which is not bound to the core. The angle of emission depends on the frequency difference between pump and signal.

In Fig. 7 we plot the total gain due to coupling to the radiation field of a high index-contrast waveguide.

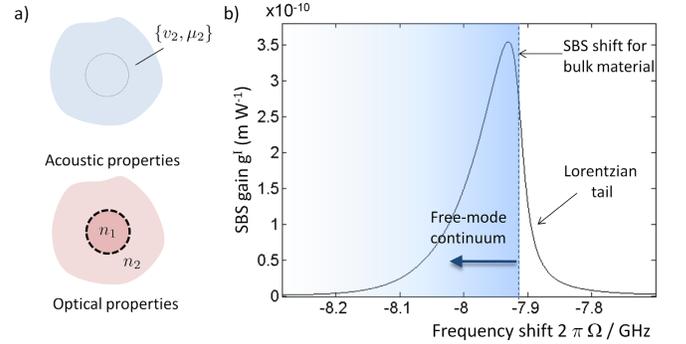

Fig. 7. a) Acoustic and optical properties for a perfectly-acoustically-matched waveguide, in which the coupling will occur the the acoustic free modes; b) Total SBS gain as a result of coupling to the continuum of non-guided acoustic modes. The SBS shift for the bulk material is at 7.9GHz.

## 6. Coupling regimes in embedded waveguides

We now examine the different SSB coupling regimes for a waveguide that has been buried in a range of solid materials. By this we aim to see how changing the acoustic properties directly affects the gain of a specific structure. We choose a rectangular waveguide with $n = 2.37$, width 4 $\mu$m and height 2 $\mu$m, embedded in an infinite matrix material with refractive index $n = 1.44$ (see Fig. 8a). This waveguide has dimensions and optical properties that are comparable to those used in recent SBS experiments [5, 8, 12]. The velocity of sound in the core is kept fixed at that of chalcogenide $v_1 = 2600\text{ms}^{-1}$, and we examine the SBS gain for different cladding materials. For the calculation of the SBS gain we used available parameters for chalcogenide glass [22]. To compute the waveguide modes, both optical and acoustic, we used a com-mercial finite element package (COMSOL); an absorbing perfectly-matched layer based on coordinate stretching [23] was implemented on the outer boundary of the mode computation in order to obtain the leaky acoustic modes.

contrast decreases the leaky modes extend further into the cladding and eventually break up into the continuum of radiation modes. In all cases a reduction in acoustic confinement is accompanied by a reduction in the SBS gain.

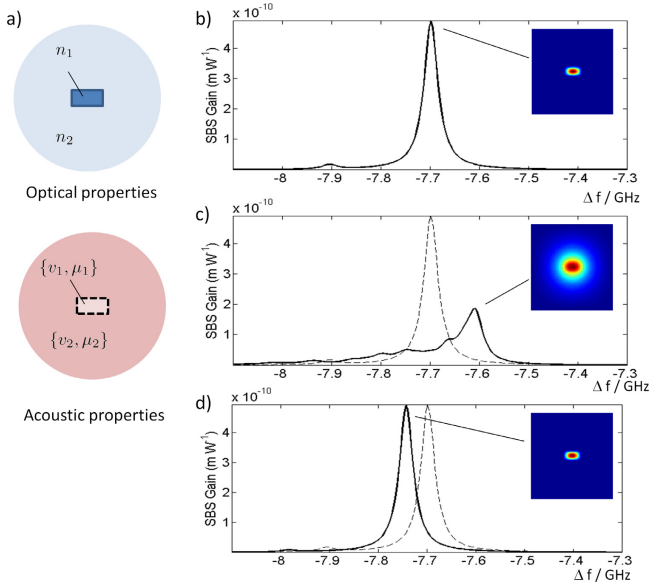

Fig. 8. a) Waveguide cross-section, showing optical and acoustic properties. b) Fundamental acoustic mode and SBS gain in the guided regime $v_1 < v_2$ c) acoustic mode and SBS gain in the radiative regime $v_1 \approx v_2$; d) acoustic mode and SBS gain in the leaky regime $v_1 > v_2$. The dotted line shows the gain in the guided case, and is included for comparison. The inset field profiles show the acoustic mode at the peak of the gain spectrum.

Keeping the optical properties constant, we examine the SBS gain in three regions: first for the case where acoustic modes are guided $v_1 < v_2$ (Fig. 8b), then for the case near the cutoff of the fundamental acoustic mode $v_1 \approx v_2$, where the coupling to the free modes can be expected to be strongest (Fig. 8c), and finally for the case where there is significant coupling to the fundamental leaky mode $v_1 >> v_2$, (Fig. 8d). For both guided and leaky modes we see significant SBS gain, owing to the strong overlap between the acoustic and optical fields. The small shift in frequency between the guided and leaky gain spectra occurs because the acoustic modes propagate at different velocities. When the acoustic properties are such that the guided acoustic mode is close to cutoff ($v_1 \approx v_2$ we see that the gain has the asymmetric shape characteristic of the coupling to free modes as shown in Fig. 7, together with a markedly reduced peak gain.

We can track the evolution of acoustic confinement in this waveguide by sweeping the ratio of acoustic velocities $v_2/v_1$ and plotting the change in peak gain (Fig. 9). We can clearly see the coupling to guided ($v_1 < v_2$) and leaky modes ($v_1 > v_2$), with the curves following very similar trajectories to those predicted by the circular waveguide model (Figs. 3a and 4a). As the velocity

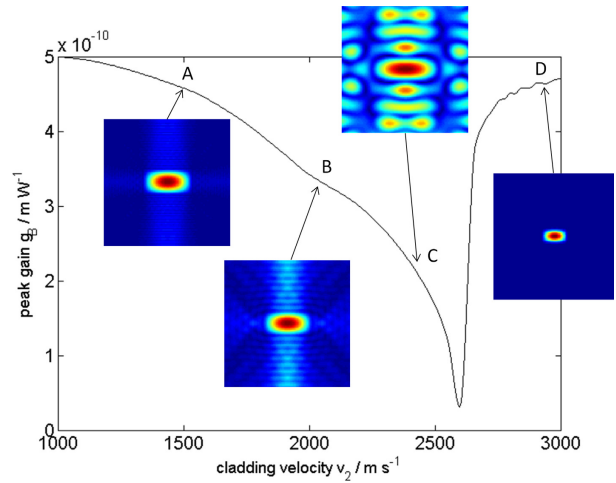

Fig. 9. SBS gain as a continuous function of the velocity ration $v_2/v_1$. The inserts show the acoustic modes in the different regimes.

## 7. Discussion

Historically, the experimental and theoretical study of SBS has focused on optical fibres, where the acoustic modes are weakly confined to the core by changing the dopant concentration [19]. The use of SBS in integrated optical waveguides is a rapidly growing field, and the conditions under which acoustic confinement can be achieved (and the SBS gain maximized) will be important for the choice of materials for core, substrate and cladding. It is noteworthy that while many materials possess refractive indices in the range of $\sim 1.5$, the range of acoustic properties of these materials can vary by a factor of five, ranging from polymers ($v \sim 1000$ ms$^{-1}$) to fused silica ($v \sim 6000$ ms$^{-1}$), with crystalline materials such as diamond possessing even higher acoustic velocities ($v \sim 12000$ ms$^{-1}$). This wide variety of material parameters gives a degree of freedom that is needed in order to design SBS-based experiments with sufficient gain.

We have seen here that there exists a class of acoustic modes that can be confined to the core in the same manner as light is confined in hollow-core waveguides. The ability to use materials with low acoustic velocities, notably polymers, as cladding materials that confine the acoustic wave opens up a number of possibilities for the use of SBS in integrated optics. The SBS gain arising from coupling to leaky acoustic modes is limited by the rate at which the mode leaks out of the core. This in turn is a function of the waveguide dimensions - as in a hollow-core optical waveguide the leakage will be reduced if the core dimensions are increased. While in-creasing the waveguide dimensions may result in other undesirable effects, such as the waveguide becoming optically multi-moded, the use of low-acoustic-velocity materials as substrates and cladding is an attractive strategy for the harnessing of SBS in high-acoustic-velocity materials such as silicon.

The findings presented here also have implications for the suppression, rather than the enhancement, of SBS. While SBS suppression can also be achieved using temperature gradients or widening of the source linewidth beyond the linewidth of SBS in the material, the gain can also be reduced by reducing the acoustic mode lifetime. This can be done by altering the dimensions of the waveguide to reduce the acoustic lifetime, as per Eq. (7), or by choosing a cladding material which has similar acoustic properties as the core. Equations (8) and (9), which give the gain for an acoustically homogenous material, place a hard limit on the amount of suppression that can be achieved in this way.

In this paper we have restricted ourselves to waveguides whose dimensions are sufficiently large that effects due to radiation pressure can be safely neglected. However it has been pointed out that for small, high-index-contrast waveguides, radiation pressure can become dominant over the electrostrictive effects that usually drive the SBS process [16]. It has been predicted that the effect of radiation pressure will lead to giant enhancement of SBS in some material systems, including Silicon suspended waveguides. Quantification of radiation pressure in waveguides that have been embedded in solids is a task that we leave for future work.

## 8. Conclusion

We have studied the effects of acoustic confinement on the SBS gain in solid-embedded waveguides, where the core may possess different acoustic properties to the cladding. We have identified three regimes of SBS gain, corresponding to coupling to acoustically guided, leaky and free modes, with the regime for any given waveguide determined by the acoustic properties of the core relative to the cladding. In both the guided and leaky acoustic wave regimes, high acoustic contrasts will result in strong confinement of the acoustic mode to the core. Furthermore, in the leaky regime high acoustic contrasts will increase the acoustic mode lifetime, allowing gain similar to that of guided acoustic modes. Finally, we have found that even when no acoustic contrast exists between the core and cladding, significant SBS gain can arise via coupling to the acoustic free modes.

### Acknowledgments

This work was supported by the Australian Research Council (ARC) through its Discovery grant (DP1096838), Center of Excellence (CUDOS, project number CE110001018) and Laureate Fellowship (Prof Ben Eggleton, Project number: FL120100029; Nonlinear optical phononics: harnessing sound and light in nonlinear nanoscale circuits) programs.

## 9. Appendix

In the following we obtain an analytic approximation for the lifetime of the acoustic leaky modes of step-index circular waveguides. The acoustic mode of a cylindrical waveguide of radius $a$ for the $m$-th azimuthal order can be written in transverse polar coordinates

$$\tilde{\rho}(r,\theta) = \begin{cases} AJ_m(k_1 r)e^{im\theta} & \text{for } r < a \\ BH_m^{(1)}(k_2 r)e^{im\theta} & \text{for } r \geq a \end{cases}, \quad (10)$$

where $k_1 = \sqrt{\Omega^2/v_1^2 - q^2}$, $k_2 = \sqrt{\Omega^2/v_2^2 - q^2}$, and $H_n^{(1)}$ is the Hankel function of the first kind. In writing such an expansion we have implicitly adopted the time dependence $\exp(-i\Omega t)$, so that the term for $r \geq a$ represents an outgoing wave. By applying boundary conditions (5) we obtain the transcendental equation for the dispersion:

$$\frac{\mu_1 k_1 a J'_m(k_1 a)}{J_m(k_1 a)} = \frac{\mu_2 k_2 a H_m^{(1)'}(k_2 a)}{H_m^{(1)}(k_2 a)}. \quad (11)$$

The equation (11) can be solved numerically to determine the frequency $\Omega$ as a function of propagation constant $q$. The leaky modes correspond to complex-valued $\Omega$ – for the time-dependence adopted here we expect $\Omega$ to possess a negative imaginary part, resulting in a mode that decays in time.

We first assume that the waveguide has sufficiently large radius that the fundamental mode is mostly contained in the core, and so has a Brillouin shift that approximates that of the core material. This implies that $q \approx \Omega/v_1$, and the angle made by the mode against the core-cladding boundary is small. We further assume that the radius is large enough that the mode is far from cutoff, and so has relatively low loss - that is, we assume that $\text{Im} V_{\text{ac}} = \Omega a\sqrt{1/v_1^2 - 1/v_2^2} \gg 1$. These two inequalities imply that $k_2 a \gg 1$, and so we obtain the relation

$$\frac{H_m^{(1)'}(k_2 a)}{H_m^{(1)}(k_2 a)} = i + \mathcal{O}\left(\frac{1}{k_2 a}\right)$$

The transcendental equation for the dispersion (11) then becomes, for the fundamental ($m = 0$) mode:

$$-\frac{J_1(k_1 a)}{J_0(k_1 a)} \approx i\frac{\mu_2 k_2}{\mu_1 k_1} \quad (12)$$

Applying a perturbation expansion of (12) using the fact that $1/(k_2 a)$ is small, we obtain

$$k_1 a = j_{0,0}\left(1 - i\frac{\mu_1}{\mu_2}\frac{1}{k_2 a}\right) + \mathcal{O}\left(\frac{1}{(k_2 a)^2}\right) \quad (13)$$

where $j_{0,0} \approx 2.405$ is the first zero of $J_0(z)$. From this equation we use the relation $\Omega = v_1\sqrt{k_1^2 + q^2}$ to obtain

$$\Omega = v_1 q \left[1 - i\frac{j_{0,0}^2 \mu_1}{q^2 a^2}\frac{1}{k_2 a} + \mathcal{O}\left(\frac{1}{(k_2 a)^2}\right)\right],$$

which gives a first approximation to the real and imaginary parts of $\Omega$:

$$\text{Re}\,\Omega = qv_1 \quad , \quad \text{Im}\,\Omega = -\frac{v_1 j_{0,0}^2 \mu_1}{\mu_2 q k_2 a^3} \qquad (14)$$

We then obtain the lifetime of the acoustic mode

$$\tau \approx -\frac{1}{\text{Im}\,\Omega} = \frac{k_2 q a^3 \mu_2}{v_1 j_{0,0}^2 \mu_1} = \frac{(\text{Re}\,\Omega)^2 a^3 \mu_2}{v_1^2 j_{00}^2 \mu_1} \sqrt{\frac{1}{v_2^2} - \frac{1}{v_1^2}} \,(15)$$